\let\useblackboard=\iftrue
%
\let\useblackboard=\iffalse
%
\newfam\black
\input harvmac.tex
\def\Title#1#2{\rightline{#1}
\ifx\answ\bigans\nopagenumbers\pageno0\vskip1in%
\baselineskip 15pt plus 1pt minus 1pt
\else
\def\listrefs{\footatend\vskip 1in\immediate\closeout\rfile\writestoppt
\baselineskip=14pt\centerline{{\bf References}}\bigskip{\frenchspacing%
\parindent=20pt\escapechar=` \input
refs.tmp\vfill\eject}\nonfrenchspacing}
\pageno1\vskip.8in\fi \centerline{\titlefont #2}\vskip .5in}

\ifx\answ\bigans\def\tcbreak#1{}\else\def\tcbreak#1{\cr&{#1}}\fi
\useblackboard
\message{If you do not have msbm (blackboard bold) fonts,}
\message{change the option at the top of the tex file.}
\font\blackboard=msbm10 scaled \magstep1
\font\blackboards=msbm7
\font\blackboardss=msbm5
\textfont\black=\blackboard
\scriptfont\black=\blackboards
\scriptscriptfont\black=\blackboardss

\else

\fi
%
\def\yboxit#1#2{\vbox{\hrule height #1 \hbox{\vrule width #1
\vbox{#2}\vrule width #1 }\hrule height #1 }}
\def\fillbox#1{\hbox to #1{\vbox to #1{\vfil}\hfil}}
\def\ybox{{\lower 1.3pt \yboxit{0.4pt}{\fillbox{8pt}}\hskip-0.2pt}}

\font\cmss=cmss10 \font\cmsss=cmss10 at 7pt
\def\IZ{\relax\ifmmode\mathchoice
{\hbox{\cmss Z\kern-.4em Z}}{\hbox{\cmss Z\kern-.4em Z}}
{\lower.9pt\hbox{\cmsss Z\kern-.4em Z}}
{\lower1.2pt\hbox{\cmsss Z\kern-.4em Z}}\else{\cmss Z\kern-.4em
Z}\fi}
\def\IR{\relax{\rm I\kern-.18em R}}
\def\comments#1{}
\def\nl{\hfill\break}
\def\BR{\IR}
\def\BZ{\IZ}
\def\p{\partial}

\def\Tr{{\rm Tr\ }}
\def\tr{{\rm tr\ }}

\def\im{{\rm Im\hskip0.1em}}

\def\CA{{\cal A}}

\def\CF{{\cal F}}
\def\CN{{\cal N}}

\def\II{\relax{I\kern-.10em I}}

\def\IIb{{\II}b}
\Title{\vbox{\baselineskip12pt
\hfill{\vbox{
\hbox{BROWN-HET-1032\hfil}
\hbox{RU-96-16\hfil}
\hbox{hep-th/9604041}}}}}
{\vbox{\centerline{D-Brane Realization of $\CN=2$  }
\vskip20pt
\centerline{Super Yang-Mills Theory in Four Dimensions}}}
\centerline{Michael R. Douglas}
\smallskip
\centerline{Dept. of Physics and Astronomy}
\centerline{Rutgers University}
\centerline{Piscataway, NJ 08855}
\centerline{\tt mrd@physics.rutgers.edu}
\bigskip
\centerline{Miao Li}
\smallskip
\centerline{Department of Physics}
\centerline{Brown University}
\centerline{Providence, RI 02912}
\centerline{\tt li@het.brown.edu}
\bigskip
\noindent
We develop and study a D-brane realization of
4D ${\cal N}=2$ super Yang-Mills theory.
It is a type IIB string theory compactified on $R^6\times K3$ and
containing parallel 7-branes.
It can also be regarded as
a subsector of Vafa's F-theory compactified on $K3\times K3$ and is
thus dual to the heterotic string on $K3\times T^2$.
We show that the one-loop prepotential in this gauge theory is
exactly equal to the interaction produced by classical closed string exchange.
A monopole configuration corresponds to an open Dirichlet 5-brane
wrapping around $K3$ with ends attached to two 7-branes.

\Date{April 1996 (revised)}
\nref\sw{N.~Seiberg and E.~Witten, Nucl. Phys. B426 (1994) 19,
hep-th/9407087.}
\nref\duality{J.~H.~Schwarz and A.~Sen, Phys. Lett. {\bf B312} (1993) 105;\nl
C. Hull and P. Townsend, Nucl. Phys. {\bf B438} (1995) 109;\nl
E. Witten, Nucl. Phys. {\bf B443} (1995) 85;\nl
A. Strominger, Nucl. Phys. {\bf B451} (1995) 96.}
\nref\joe{J. Polchinski, ``Dirichlet-Branes and Ramond-Ramond Charges,''
hep-th/9510017.}
\nref\AF{P.~C.~Argyres and A.~E.~Faraggi, hep-th/9411057.}
\nref\KLTY{A.~Klemm, W.~Lerche, S.~Yankielowicz and S.~Theisen,
hep-th/9411048 and hep-th/9412158.}
\nref\greens{B. Greene, A. Shapere, C. Vafa and S.-T. Yau,
Nucl. Phys. B337 (1990) 1;\nl
G. Gibbons, M.B. Green and M.J. Perry, hep-th/9511080.}
\nref\vafaf{C.~Vafa, ``Evidence for F-theory,'' hep-th/9602022.}
\nref\bp{C. Bachas and M. Porrati, Phys. Lett. B296 (1992) 77,
hep-th/9209032.}
\nref\hm{J. Harvey and G. Moore, ``Algebras, BPS States, and Strings,''
hep-th/9510182.}
\nref\bound{E. Witten, ``Bound States of Strings and P-branes,''
hep-th/9510135.}
\nref\cs{E. Cremmer and J. Scherk, Nucl. Phys. B72 (1974) 117.}
\nref\andy{A. Strominger, ``Open P-Branes,''  hep-th/9512059.}
\nref\dl{M. Douglas and M. Li, unpublished, and talk given by
M. Douglas on Jan. 25 at ITP, Santa Barbara.}
\nref\cumrun{S. Kachru, A. Klemm, W. Lerche, P. Mayr and
C. Vafa, ``Nonperturbative Results on The Point Particle Limit
of N=2 String Compactifications,'' hep-th/9508155.}
\nref\miao{M. Li, ``Boundary States of D-Branes and Dy-Strings,''
hep-th/9510161, to appear in Nucl. Phys. B.}
\nref\mike{M. Douglas, ``Branes within Branes,'' hep-th/9512077.}
\nref\ck{C. G. Callan and I. R. Klebanov, ``D-Brane Boundary State
Dynamics,'' hep-th/9511173.}
\nref\sch{C. Schmidhuber, ``D-Brane Actions,''  hep-th/9601003.}
\nref\das{S. P. de Alwis and K. Sato, ``D Strings and F Strings
from String Loops,''  hep-th/9601167.}
\nref\bvs{M. Bershadsky, C. Vafa and V. Sadov, ``D-branes and
Topological Field Theories,'' hep-th/9511222.}
\nref\bps{E. B. Bogomolny, Sov. J. Nucl. Phys. 23 (1976) 435;\nl
M. K. Prasad and C. M. Sommerfield, Phys. Rev. Lett. 35 (1975) 760.}
\nref\mlpa{M. R. Douglas and M. Li, to appear.}
\newsec{Introduction}
Much progress has been made in understanding dualities in
supersymmetric Yang-Mills theories (SYM) and string theory.
\refs{\sw,\duality}\ %
It is especially striking that field theory duality
can be viewed as consequence of string dualities. As one example,
the self-duality of ${\cal N}=4$ SYM can be understood as
a consequence of self-duality of  ${\cal N}=4$ heterotic string
compactified on $R^4\times T^6$. If one places parallel 3-branes
\joe\ in the type IIB theory on $R^{10}$, one obtains a ${\cal N}=4$
4D gauge theory as the world-volume theory \bound, and the
self-duality of SYM is a consequence of the self-duality of the
type \IIb\ theory in ten dimensions.
Dyon solutions of the world-volume theory have an interesting 10D spacetime
interpretation, explored in \refs{\andy,\dl}. As another example,
the celebrated Seiberg-Witten solution of ${\cal N}=2$ SYM
can be found using type \II-heterotic duality in four dimensions
\cumrun. Given that the ${\cal N}=4$ SYM can be realized
by D-branes, naturally one would like to know whether
${\cal N}=2$ SYM can also be realized as a world-volume
theory of D-branes. This paper will give such a construction.

We shall first discuss various D-branes with some compact dimensions
wrapping around holomorphic cycles in $K3$ in section 2. Detailed analysis
is given to a realization using parallel 7-branes in section 3, where
we also discuss its relation with F-theory.
In section 4, we discuss the one-loop prepotential, while
in section 5, we study the equivalence of monopoles in the gauge
theory and D-branes, and show that in the 7-brane construction
a monopole corresponds to an open 5-brane wrapping around $K3$.

\newsec{Branes in $R^6\times K3$}

Compactification of the type IIB theory on $R^6\times K3$
produces a chiral six dimensional theory with $(0,2)$
SUSY, reducing to $\CN=4$ SUSY in 4D.
Introducing D-branes will break at least half of this supersymmetry.
If it is half, the world-volume theory will be a $(0,1)$
SYM in 6d.  If $N$ D-branes fill four of these six dimensions,
their world volume theory will be $\CN=2$ SYM
in 4d with gauge group $U(N)$,
so this is a natural setting for our project.

We consider the type IIB theory, with D$p$-branes for all odd $p$.
An $\CN=2$ SYM in 4d could be produced by wrapping 7-branes
around the entire $K3$, wrapping 5-branes around a holomorphic curve,
or placing 3-branes at a point in $K3$.

The $4d$ matter content is found by reducing the world-volume theory on $K3$.
Let $\Sigma$ be the $2n$-cycle about which the D-branes are wrapped;
then the $4d$ matter is a sigma model whose target is the moduli space
of vacua of a twisted $U(N)$ gauge theory on $\Sigma$ with scalars in the
normal bundle to $\Sigma$.\ \bvs\ %

For $\Sigma=K3$, the moduli space is the space of flat connections on $K3$,
which is trivial.  The 4d theory is pure SYM.

In the case $\Sigma = {\rm pt}$,
the 4d matter is a sigma model on $K3\times V$ where $V$ is the
adjoint representation of $U(N)$.  Such sigma models can be gauged
(preserving $\CN=2$ SUSY) if $V$ is a real representation of the gauge
group.  This theory has the matter content of $\CN=4$ SYM, broken to
$\CN=2$ by the curvature of the K3.

Finally, for $\Sigma$ a holomorphic curve,
the 4d matter fields parameterize the moduli space of solutions to a Hitchin
system \bvs, which can produce other charged matter.  We will not discuss
this interesting case here, except to mention that for genus zero, one again
obtains pure SYM.

\newsec{$7$-branes and F-theory}

We proceed to study a system of $N$ $7$-branes wrapped around K3.
Since the space transverse to a $7$-brane is two-dimensional, it has
a deficit angle at infinity, calculated to be $\pi/6$. \greens\ %
Furthermore, the dilaton and axion are non-constant.
Non-compact multi-brane solutions can be found with $N\le 12$ branes.

Another option is to take $N=24$ for which the total curvature is $4\pi$
and the transverse space closes into an $S^2$.
As pointed out by Vafa \vafaf, to get the total monodromy of $\tau$ around the
$24$ branes to vanish, we are forced to take not just D-branes but in
addition ``$(p,q)$''-branes, $SL(2,\BZ)$ images of the basic
D-brane.

Vafa has proposed that this system of $7$-branes on $\BR^8\times S^2$
is a strong coupling dual of the heterotic string on $\BR^8\times T^2$.
It can also be regarded as a compactification of ``$12$-dimensional
F-theory'' on $K3$.
Thus, the theory we are considering is
F-theory on $K3\times K3$,
a strong coupling dual of the heterotic string on $K3\times T^2$.

An important point is that the dilaton and axion in the type \II\ theory
are determined by the equations of motion, and are no longer moduli.
The duality transformation on the low energy Lagrangians
relates the parameters in the dual theories (here $\alpha'=1$) as
\eqn\duality{\eqalign{
	{1\over\lambda_8^2} &= {1\over V_{\II}(S^2)^2}
			= e^{-2\phi_h} V_h(T^2)\cr
	M_{\II} &= V_{\II}(S^2)^{1/2}\ M_h.
}}
The weak coupling limit is small $S^2$.
Here $M_{\II}$ and $M_h$ are generic mass scales in eight dimensions in
the two theories.  Note that a heterotic state with $M\sim 1$ is related
to a type \II\ state with $M\sim V(S^2)^{1/2}$, for example a string stretching
between $7$-branes.

Up to $18$ of the $7$-branes can be Dirichlet or $(1,0)$ branes.
Although our subsequent analysis will consider only this sector,
let us first make a few comments about $(p,q)$ $7$-branes.
First, since there is no local way to determine $(p,q)$,
each $7$-brane will come with a $U(1)$ gauge symmetry,
and the total gauge symmetry from $7$-branes is $U(1)^{24}$.
Duality with the heterotic string
will require this $U(1)^{24}$ to be broken to $U(1)^{20}$,
but this must be due to global effects.

Encircling a Dirichlet $7$-brane induces the $SL(2,\BZ)$ monodromy $T$ on
$SL(2,\BZ)$ doublet fields, for example
$(B^{(2)},C^{(2)})|_{\theta=2\pi} = (B^{(2)},C^{(2)}+B^{(2)})|_{\theta=0}$.
One should keep in mind that this monodromy takes states to equivalent states
described in different conventions.
For example, taking a $(p,q)$ string around the D$7$-brane produces an
object with the same tension, because both $(p,q)$ and the dilaton-axion
transform.

We could define a $(p,q)$ $7$-brane in two ways.
One way is to start with the Dirichlet $7$-brane and apply a general
$SL(2,\BZ)$ transformation $g=\left(\matrix{ p & r\cr q & s }\right)$.
The resulting solution depends on the integers $(p,q,r,s)$ (with $ps-qr=1$).
However, before accepting the conclusion that a
$(p,q,r,s)$ brane is different from a $(p,q,r',s')$ brane,
one should show that this is a physical difference between the branes
and not just the backgrounds.
In particular, the $SL(2,\BZ)$ monodromy produced by encircling the
brane, $g T g^{-1}$, depends only on $(p,q)$.

Another definition is that we label the $7$-brane by the type of
objects which can end on it.  Consider a $7$-brane which is
an allowed endpoint for $(p,q)$ dyonic strings, and $(t,u)$ dyonic
$5$-branes.  Such an object would necessarily have world-volume couplings
\eqn\wpqg{\int d^8x\left( (t\tilde{B}^{(6)}+uC^{(6)})\wedge F
+(pB^{(2)}+qC^{(2)} - {F})^2\right),}
where $\tilde{B}^{(6)}$ is the dual of $B^{(2)}$ \andy.

However, such couplings are only sensible if the fields involved are
single-valued.  This requires $(t,u)$ to be $n(p,-q)$ (for $n$ integer),
and thus the $1$-brane and $5$-brane to be mutually local.
If we assume that the gauge field is invariant under $SL(2,\BZ)$,
transforming the known D-brane couplings produces
\eqn\wpq{\int d^8x\left( (pC^{(6)}-q\tilde{B}^{(6)})\wedge F
+(pB^{(2)}+qC^{(2)} - {F})^2\right),}
in agreement with the above considerations with $n=1$.

We conclude that the evidence is consistent with a $7$-brane
labelled by the two integers $(p,q)$ with $p\ge 0$.
(The overall sign of $(p,q)$ can be set by using the freedom
to take $F\rightarrow -F$.)

As we commented, duality with the heterotic string requires
the $U(1)^{24}$ of the $7$-branes to be broken to $U(1)^{20}$.
A sign of this can be seen by considering
the neighborhood of a group of $N$ mutually local branes,
with broken $U(N)$ gauge symmetry.
There, the couplings \wpq\ will allow the locally single-valued
tensor $pB^{(2)}+qC^{(2)}$ to `eat' the diagonal $U(1)$ gauge boson
(the Cremmer-Scherk mechanism \cs), breaking this $U(1)$.
We will find in section 4 that this breaking will
solve a paradox associated with loop effects.
Understanding the global reduction of the gauge group is more subtle
and we will discuss this in \mlpa.

After further compactification on K3,
the effective four dimensional gauge coupling
constant is $g^2\sim V(S^2)^2/V(K3)$, where
$V(K3)$ is the volume of $K3$, and the small
volume limit of $K3$ is the strong coupling limit of the
four dimensional effective theory. Of course this is classical
and in this ${\cal N}=2$ theory, the coupling constant will be renormalized.

\newsec{Renormalization}

The string coupling constant for type II theory
on the background $R^6\times K3$
does not receive renormalization, thanks to its $(0,2)$ supersymmetry.
However, the D-brane world-volume ${\cal N}=2$, $d=4$ SYM
has a beta function, non-zero at one-loop.
It can be expressed as a quantum correction to the prepotential of
an effective theory valid in the case of gauge symmetry breaking to $U(1)^N$
\refs{\AF,\KLTY}:
\eqn\Foneloop{\CF_1 = {i\over 4\pi}
\sum_{i<j} (A_i-A_j)^2 \log {(A_i-A_j)^2 \over e^3\Lambda^2}.}
Here $(A_i,W_i)$ are the $\CN=1$ chiral and vector superfields which
make up an $\CN=2$ vector multiplet.
In terms of the D-brane configuration,
the scalar component of $A_i$ is the position $X^4_i+iX^5_i$ of the $i$'th
D-brane.

The open string diagram relevant for the beta function of the world-volume
theory is the `W-boson' loop diagram,
an annulus with one boundary on D-brane $i$ and another on
D-brane $j$.  (There is no matter charged under a single $U(1)$).
The term
\eqn\lagp{
\im \int d^2\theta\ \p_i\p_j \CF_1(A) W_i W_j = {1\over 32\pi^2 }
\sum_{i<j} \log {|a_i-a_j|^2} (F_i-F_j)^2 
}
in the effective Lagrangian can be seen by doing this calculation
in a constant background field $F$.
A priori, we would expect this result to be valid for $|a_i-a_j|^2<<\alpha'$,
while for larger separations
the massive open string states could
give equally important contributions.
As it will turn out, \lagp\ is exact.

The annulus diagram has a dual interpretation as a closed string exchange
between the branes.  Thus the one-loop prepotential can be obtained by
a purely classical computation: we need to find the closed string source
corresponding to the constant background field $F$; then the amplitude will
be a weighted sum of free particle Green's functions.

In the large separation limit $|a_i-a_j|^2>>\alpha'$, it will be dominated by
massless closed string exchange.
It is amusing to see that the prepotential will also be logarithmic in
this regime, simply because it is proportional to the free
massless Green's function
$G(X_i-X_j)$ in two transverse dimensions.

In section 3
we considered the couplings to massless closed string fields
present before compactification on K3.  Their couplings respect
$\CN=4$, $d=4$ supersymmetry and should not lead to a beta function.
The new fields on K3 are zero modes produced by using harmonic forms.
Besides the volume form, there are $19$ anti-self-dual and $3$ self-dual
two-forms.
Reduction of $C^{(4+)}$ on these produces tensor fields with couplings
on the effective $3$-brane volume of the form
\eqn\twvact{
Q\int d^4x\ \tilde C_i^{(2)}\wedge F.
}
We would expect a 7-brane to couple with equal strength to each, leading to
the result $\CA\sim(19-3)Q^2 \log X$.  The charge $Q$ might in principle
depend on the type of $(p,q)$ brane.

To get the exact result and check the normalization, we now do the one-loop
open string computation for the orbifold $T^4/Z_2$.
We consider two D7-branes on each with a constant
electric field $E_i=F^i_{01}$.
Quantization of open strings stretched between these branes
is done as in \joe, except that two longitudinal
coordinates are quantized in the background fields $E_i$.
This was done in \bp, and we shall make use of their results. Let $L_0$
denote the open string Hamiltonian.
The orbifold projection $R$ acts as
\eqn\orbprot{
R X^i R^{-1}=-X^i, \quad R \psi^i R^{-1}=-\psi^i.
}
Equivalently, writing $R=R_bR_f$ where $R_b$ acts only on bosons and
$R_f$ only on fermions, $R_f=(-1)^{F_2}$ where
$F_1$ is fermion number associated to $R^6$
and $F_2$ fermion number associated to $T^4$.
We also use the convention
$(-1)^{F_1}|0\rangle_{NS}=-|0\rangle_{NS}$ and
$(-1)^{F_2}|0\rangle_{NS}=|0\rangle_{NS}$.

The one-loop amplitude is then
\eqn\oneloop{
\CA = 2\int {dt\over 2t}\tr e^{2\pi iF}\ %
{(1+(-1)^F)\over 2} {(1+R)\over 2} q^{L_0}
}
where $q=\exp(-2\pi t)$ and $e^{2\pi iF}=-1$ in the Ramond sector.

Let us separate the untwisted and twisted closed string
sectors by regrouping the projections:
\eqn\trick{{(1+(-1)^F)\over 2} {(1+R)\over 2}
= {(1+(-1)^F)\over 4} +
R_b{((-1)^{F_1}+(-1)^{F_2})\over 4}.}
The untwisted sector
(the first term on the right) produces half of the one-loop result on $T^4$.
It can be obtained by combining results
of \joe\ and \bp:
\eqn\calc{\eqalign{
\CA = &{V_4\over 4} \int {dt\over t}(2\pi t)^{-2}
e^{-tX^2/2\pi\alpha'}\sum_p q^{\alpha' p^2}
	\prod_{n\ge 1}(1-q^{n})^{-8} f_B(q,E_1,E_2) \cr
&[-16\prod_{n\ge 1}(1+q^{n})^{8}
		{\Theta[{1-2i\epsilon\atop 0}]\over\Theta[{1\atop 0}]}
+q^{-1/2}\prod_{n\ge 1}(1+q^{n-1/2})^{8}
		{\Theta[{2i\epsilon\atop 0}]\over\Theta[{0\atop 0}]}\cr
&+q^{-1/2}\prod_{n\ge 1}(1-q^{n-1/2})^{8}
		{\Theta[{2i\epsilon\atop 1}]\over\Theta[{0\atop 1}]}
]}}
with
$$\pi\epsilon =\tanh^{-1}\pi E_1-\tanh^{-1}\pi E_2$$
and
\eqn\calcdef{\eqalign{
f_B(q,E_1,E_2) &= {\pi(E_1+E_2)\over t}{q^{\epsilon^2/2}\over q^{-i\epsilon/2
}-q^{i\epsilon/2}}
\prod_{n\ge 1}{(1-q^{n})^2\over (1-q^{n+i\epsilon})(1-q^{n-i\epsilon})}\cr
{\Theta[{a\atop b}]\over\eta} &= q^{{1\over 8}a^2-{1\over 24}}
	\prod_{n\ge 1}(1+e^{i\pi b}q^{n+(a-1)/2})(1+e^{i\pi b}q^{n-(a-1)/2}).
}}
The sum $\sum_p$ in \calc\ is over internal momenta associated to torus
$T^4$.
Expanding to $O(E_1E_2)$, it is easy to see that at both limits
$t\rightarrow 0$ and $t\rightarrow\infty$, all terms cancel.
Thus by general properties of modular functions the integrand must vanish.

The nonvanishing part to quadratic order in $E$ comes from the second
term, the twisted sector, and solely from the (open string) NS sector: it is
\eqn\onel{\eqalign{
F_1={(E_1-E_2)^2\over 32\pi^2}\int {dt\over t}
&e^{-t{X^2/2\pi\alpha'}}\ \left[\sum_{n=1}{q^{n-1}(1+q^{2n-1})
\over (1-q^{2n-1})^2}\right]\cr
&\tr_{T^4 \rm bosons}\left( R_bq^{L_0}\right)
 \prod_{n=1}(1-q^n)^{-4}(1-q^{2n-1})^4.
}}
Since this comes from the twisted sectors, it is
independent of internal momentum, therefore independent of
the size and shape of $T^4/Z_2$.

It is easy to see that the bosonic
trace is $\tr R_bq^{L_0}=\prod_{n=1}(1+q^n)^{-4}$.
Furthermore, the expression in square brackets can be simplified
(by expanding $1/(1-q^{2n-1})^2$, summing over $n$ and using theta
function expansions) to
$\prod_{n=1}(1-q^n)^{4}(1+q^n)^{8}$.  Finally,
$\prod_{n=1}(1+q^n)(1-q^{2n-1})=1$ and the massive contributions all cancel:
\eqn\onell{F_1={(E_1-E_2)^2\over 32\pi^2}\int {dt\over t}
\exp(-t{X^2\over 2\pi\alpha'}).}
The integral is divergent at $t\rightarrow 0$, which means closed string
proper time $1/t\rightarrow\infty$.
This is the usual space-time IR divergence of the two-dimensional
bosonic Green's function given sources of non-zero total charge.

One response to this divergence is simply to work with $N\le 12$ branes
and a non-compact space, identify it as the signal of growing fields
at infinity, and ignore it.  This is not a very satisfying response however
as it means that a subsector of a larger theory would couple strongly
to subsectors far away in $X$.

Let us consider
a group of mutually local $7$-branes, far away from the rest of the
system.  For such a group, the total source in \twvact\ will
be the field strength $F$ in the diagonal $U(1)$.
As we argued in section 3, this will be broken.  Thus the total source
will be zero and the IR divergence produced by this group will cancel.

In the theory with $N=24$ branes, the two dimensions are compact,
and consistency of the equations of motion requires the total charge
$\sum_i Q_i F_i$ and thus the IR divergence to cancel.
Not having the contributions from all $(p,q)$ branes under control,
we cannot prove that this works, but given that
there are $4$ broken $U(1)$'s for which $F$ is guaranteed
to be zero, one can hope that the total source $Q_i$
couples to one of these.
If so, cancellation of the total IR divergence will imply that the
$\alpha'$ in \onell\ cancels out of the final result.
Thus, in a subsector,
the scale $\Lambda$ in $\Foneloop$ will be set by global effects.

The final result is that the one-loop
prepotential has precisely the form \lagp.
We can think of it as either coming only from the lightest
(no oscillator excitation and $m=|X|/\alpha'$) open string
loops, or the massless closed string exchange.

There is a simple argument for the absence of higher open string
corrections to these results.
In similar calculations such as \hm,
it was seen that the one-loop result
\Foneloop\ only received contributions from BPS states;
the contributions of non-BPS states cancelled.
In the present problem, the central charge of a particle state is
determined by its $U(1)^{20}$ charge, which is determined just by
the endpoints of the string.  Thus any oscillator contribution to
the mass will raise it above the BPS bound.

Although the fact that an amplitude involving a finite number of states can
be dual in the world-sheet sense may be surprising, this is a common
feature of two-dimensional string theories.  Here we see that it can be
true of a two-dimensional subsector of a physical string theory.

There are additional $E_1^2$ and $E_2^2$ terms in the result, which could
have been predicted by the absence of states charged under the overall $U(1)$.
These must come from exchange of the fields $\tilde B^{(0)}$ and
$\tilde C^{(0)}$, with both a tadpole and a source $F^2$.
IR divergences must cancel for all three fields separately;
we believe that enforcing the condition $0=\int \Tr F^2-\tr R^2$
will be necessary for this.

Further corrections to the prepotential will be due to instantons,
which in this picture are 3-branes wound around $K3$.

\newsec{Monopoles}
\nref\green{M. B. Green and M. Gutperle, ``Comments on Three-Branes,''
hep-th/9602077.}
\nref\tseytlin{A.A. Tseytlin, ``Self-duality of Born-Infeld action and
Dirichlet 3-brane of type IIB superstring theory,'' hep-th/9602064.}

Electric charged excitations on the effective 3-branes constructed
in the last section are open strings with their ends attached to
branes.  Is there a D-brane interpretation of the gauge theory monopoles?

It was shown in \andy\ that a Dirichlet (p-2)-brane can
end on Dirichlet p-branes.
The simplest case is an open D-string with ends attached to parallel 3-branes
in $\BR^{10}$, and each end appears as a magnetic charge in its 3-brane.

It was argued in \dl\ that indeed
this configuration corresponds to the monopole solution in the
3-brane world-volume theory.\footnote*{
After this section was completed, the papers
\refs{\green,\tseytlin}\ appeared,
in which this interpretation was also explored.}
The simplest argument is that the
$SL(2,\BZ)$ duality of IIb string theory includes the $SL(2,\BZ)$
self-duality of the $3$-branes, and the fundamental open string and D1-brane
stretched between the 3-branes form a doublet.  Thus this duality becomes the
$SL(2,\BZ)$ duality of the $\CN=4$ SYM theory on the 3-branes, with the
doublet becoming the $W$-boson and monopole.  Likewise, a dyon solution
coincides with an open dy-string with ends attached to 3-branes.

To explore the relation between the two descriptions,
we take $\CN=4$ SYM with adjoint fields $A_\mu$ and $X^I$, $4\le I\le 9$,
and consider a vacuum with the two 3-branes at $X^4=\pm c/2$,
the other $X^I=0$.
We take $\alpha'=1$.
In a one monopole solution (written using a single gauge patch) we have \bps
$$X^4={1\over 2r}(1-cr\hbox{coth}cr)\sigma_3,$$
so asymptotically $X_4=-c/2\sigma_3$, but in the core,
the distance approaches zero, reaching zero at $r=0$.

Far outside the
core of the monopole, the two 3-branes are separated in transverse
space, while they become indistinguishable at the core.
The geometry is similar to that of the $D1$-brane and in this description
we can think of the lightness of the $W$'s and the
gauge symmetry restoration at the core as coming from
the possibility that an open fundamental string can bend
to touch the $D1$-brane, forming a dy-string there which costs zero
energy.  This picture also suggests that
the RR fields produced by the monopole and $1$-brane have the same structure
in space, which is not hard to verify.

Amusingly, in this non-singular gauge, the question of ``which'' $3$-brane is
at which point $X^4=\pm c/2$ at infinity depends on which direction one goes
out to infinity!  Of course this is not a gauge-invariant statement and
after introducing two gauge patches,
the non-trivial behavior takes place only in the core.

Extending this analysis to parallel 7-branes, we shall find that
monopoles are open Dirichlet 5-branes wrapping around $K3$, they appear
as open strings in $R^6$. Dyons must be interpreted as bound
states of open strings and open Dirichlet 5-branes.
The lacking of self-duality in Seiberg-Witten
theory has one obvious origin in D-brane realization: The
electric charged states are open strings, while magnetic charged
states are open 5-branes.

For simplicity, we consider the case of two parallel 7-branes. The
effective world-volume theory is a $U(2)$ SYM. There are two adjoint
scalars which we denote by $\phi_4$ and $\phi_5$, since they arise
from fluctuations of 7-branes in transverse directions $x^4$ and $x^5$. These
are $2\times 2$ Hermitian matrices. There is a potential
$-[\phi_4,\phi_5]^2$ in the action, so a classical minimum
is given by mutually commuting $\phi_4$ and $\phi_5$. Then these
two matrices can be diagonalized simultaneously, and the eigenvalues
will be transverse locations of the two branes in transverse space.
Now a classical monopole solution
breaks half of the world-volume supersymmetry.

We now show that in for parallel 7-branes,
this solution corresponds to one open 5-brane connecting two 7-branes.
To do this, we recall the equation for massless RR--D-brane coupings  given in
\mike, which generalizes
the result of \miao\foot{For related considerations, see \refs{\ck\sch\das}.}:
\eqn\aneq{d^* H=\tr e^{F+D_\mu\phi_i dx^\mu\wedge b^i*} J,}
where $H$ is a sum of the field strength of all R-R tensor fields,
$F$ is the field strength of gauge field as a two-form. The term
$D_\mu\phi_i dx^\mu\wedge b^i$ is the T-dual version of $F$. The symbol $b^i$
stands for $^*dx^{i*}$, when it acts on a form its effect is to eliminate
the factor $dx^i$ in that form, otherwise the result is zero.

For a single D-brane, $J$ is just the world-volume current
$J=\delta(x^i-x^i_0)dx^0\wedge\dots \wedge dx^p$. For a system of
multi-branes, $J$ needs some elaboration. For two parallel D-branes,
the single delta-function is replaced by a sum of two delta-functions.
This sum can be understood as coming from action of the zero mode
of the world-sheet fields $x^i$: A dual Wilson line $\tr\exp(\oint \phi_i\p_n
x^id\sigma)$ is to be inserted as a boundary term of the open
string world-sheet. For constant and simultaneously diagonalized
$\phi_i$, the result is just $\sum_a\exp(i\phi_i^ap^i)$, a sum
of shifting operators in the transverse space. This yields a sum
of delta-functions when it acts on a single delta-function, each
delta-function represents a D-brane. Now if $\phi_i$ is a function
of $x^\mu$, the effect of the zero mode of $\p_nx^i$ will be the same as
the above, as long as $\phi_i$ is slowly varying. So \aneq\ applies
to the monopole solution if the separation $c$ is much smaller than
the string scale, since the core size is proportional to $1/c$.

For two 7-branes, $^*J$ is a two-form in the transverse space. Expanding
the R.H.S. of \aneq, we obtain the first term $^*J$ which provides
a source for $^*H^{(9)}$. This is the standard result, with charge
given by
$$\int_{S^1}\hskip2pt ^*H^{(9)},$$
where $S^1$ encircles 7-branes. The next term upon expanding \aneq\
is proportional to $\tr F_{ij}D_k\phi_4 dx^i\wedge dx^j\wedge
dx^k\wedge dx^5$, providing a source for $^*H^{(7)}$, the
electric field corresponding to a 5-brane. Using the delta-functions,
it is easy to see that this electric field is confined in between
the two 7-branes, but the explicit form is complicated. To compute
the charge, it is better to consider the average of the charge
\eqn\elec{{1\over c}\int^{c/2}_{-c/2} dx^4\int_{S^3}\hskip2pt ^*H^{(7)}}
here $S^3$ encircles the 5-brane. This integral then becomes
$${1\over c}\int dx^4d ^*H^{(7)},$$
Make use of eq.\aneq\ and the delta-functions,
the integral is reduced to a 3D integral on the world-volume. We then use the
self-duality of the monopole solution to arrive at
\eqn\resl{\eqalign{&{1\over c}\int^{c/2}_{-c/2} dx^4\int_{S^3}
\hskip2pt ^*H^{(7)}\cr
&={2\over c}\int d^3x\tr\sum_i(D_i\phi)^2=4\pi,}}
the result is independent of $c$.
It is obvious that this 5-brane is an  open one with
four compact dimensions $K3$, as $S^3$ is embedded in $R^6$.
As a further check of this identification, note that the mass of
the monopole is proportional to $c/g^2\sim cV(K3)/\lambda$, this
matches the fact that the tension of the open 5-brane is proportional
to $1/\lambda$ and its volume is $cV(K3)$.
There are no further terms from the R.H.S. of \aneq, so no other
R-R tensor fields are generated. In the T-dual picture in the type IIA
theory, open 5-branes described here become open 6-branes wrapping around
$S^1\times K3$.

One may ask what happens to other open branes with ends on 7-branes.
For example one may imagine an open 3-brane with two compact dimensions
wrapping around a holomorphic curve in $K3$. To generate such a
configuration, apparently the internal fields $A_a$ would have to
be switched on. From eq.\aneq, one easily see that $A_a$ must be
nonvanishing along the holomorphic curve. Similarly, an open one-brane
demands all four components  $A_a$ be switched on. From the point of view
of the effective four dimensional theory on the world-volume of 7-branes,
these configurations cost much energy.

\newsec{Discussion}

We have exhibited in this paper D-brane realizations of
$\CN=2$ super Yang-Mills theory in four dimensions.
Using a realization as 7-branes in the type IIB theory on $R^6\times K3$,
we computed the one-loop prepotential and exhibited monopoles as
5-branes.
The prepotential turned out to have a dual interpretation as the
classical force between gauge field excitations on the branes, mediated
by closed string exchange.
It would be very interesting to obtain Seiberg and Witten's solution
(and its $SU(N)$ generalization by Argyres et. al. and Klemm at. al.)
in this picture.

The theory fits naturally into a compactification of
`F-theory' on $K3\times K3$ and is
dual to the heterotic string on $K3\times T^2$.
We believe that with further work to make the map between the theories
more explicit,
it will be possible to relate the one-loop results here
to exact results for the one-loop prepotential in the heterotic theory.

\medskip
\noindent {\bf Acknowledgments}

M.R.D. would like to thank T. Banks, W. Lerche, G. Moore, J. Schwarz
and N. Seiberg for useful discussions and correspondance, 

and the ITP Santa Barbara for its
hospitality.
His work was supported by NSF grant PHY-9157016 and
DOE grant DE-FG05-90ER40559.
The work of M.L. was supported by DOE grant DE-FG02-91ER40688-Task A.

\medskip
\nref\lerche{A. Klemm, W. Lerche, P. Mayr, C. Vafa and N. Warner,
``Self-Dual Strings and N=2 Supersymmetric Field Theory,'' hep-th/9604034.}
After completion of this work, we received the paper \lerche,
which also discusses D-brane realizations of $\CN=2$ gauge theories.

\listrefs

\end